\documentclass[aps,pre,twocolumn,groupedaddress,showpacs]{revtex4}
\usepackage{amssymb}
\usepackage{graphicx,color}
\definecolor{r}{rgb}{1,0,0}
\definecolor{g}{rgb}{0,1,0}
\definecolor{b}{rgb}{0,0,1}

\begin{document}


\title{Effect of hydrogel particle additives on water-accessible pore structure of sandy soils: A custom pressure plate apparatus and capillary bundle model}


\author{Y. Wei$^{1,2}$, D. J. Durian$^1$}
\affiliation{$^{1}$Department of Physics and Astronomy, University of Pennsylvania, Philadelphia, PA 19104-6396, USA}
\affiliation{$^{2}$Complex Assemblies of Soft Matter, CNRS-Rhodia-UPenn UMI 3254, Bristol, PA 19007-3624, USA}


\date{\today}

\begin{abstract}
To probe the effects of hydrogel particle additives on the water-accessible pore structure of sandy soils, we introduce a custom pressure plate method in which the volume of water expelled from a wet granular packing is measured as a function of applied pressure.  Using a capillary bundle model, we show that the differential change in retained water per pressure increment is directly related to the cumulative cross-sectional area distribution $f(r)$ of the water-accessible pores with radii less than $r$.  This is validated by measurements of water expelled from a model sandy soil composed of 2~mm diameter glass beads.  In particular, the expelled water is found to depend dramatically on sample height and that analysis using the capillary bundle model gives the same pore size distribution for all samples.  The distribution is found to be approximately log-normal, and the total cross-sectional area fraction of the accessible pore space is found to be $f_0=0.34$.  We then report on how the pore distribution and total water-accessible area fraction are affected by superabsorbent hydrogel particle additives, uniformly mixed into a fixed-height sample at varying concentrations.   Under both fixed volume and free swelling conditions, the total area fraction of water-accessible pore space in a packing decreases exponentially as the gel concentration increases.  The size distribution of the pores is significantly modified by the swollen hydrogel particles, such that large pores are clogged while small pores are formed.
\end{abstract}

%


\maketitle





Capillary storage of water is an important property that contributes to the plant water availability in soils, especially in sandy soils.  When the capillary forces are strong compared to gravity, the rain water is trapped inside the pores and may be used to support the growth of plants.  Since capillary forces depend on pore size, the amount of capillary water inside a sandy soil is tightly linked to its pore structure.  An improvement in the water retention of a sandy soil usually couples with a change in the water-accessible pore structure.  As a popular soil additive, superabsorbent hydrogel particles have been proven to efficiently enhance water retention of sandy soils by swelling and hence locking water inside themselves \cite{Azzam83, Johnson84, Kazanskii92, Bouranis95, Singh, Buchholz, Bhardwaj07, Andry09, Bai10}.  However, it is yet to be clarified the extent to which the improvement is also due to the modification of the water-accessible pore structure caused by the presence of the hydrogel particles.

A standard way to determine the soil water retention is to use a pressure plate apparatus, introduced by Richards in 1940's \cite{Richards41, Richards43}.  The basic idea is to measure the amount of solution expelled from a wet soil under a given pressure head.  For this, a water-saturated soil sample is placed on an extraction chamber whose bottom is embedded with a wet porous plate.  When an extra gas pressure is applied, the wet porous plate allows soil water to flow out but prevents the escape of the compressed gas in the chamber.  In the past several decades, this apparatus has been widely used in soil research to determine the so-called soil-water characteristic curve (SWCC), $\theta (P)$ vs $P$, which is defined as the ratio of the water volume $V_{water}$ retained in soil under a given suction pressure $P$ to the initial volume of the dry soil $V_{soil}$:
\begin{equation}
   \theta (P) = \frac{V_{water}(P)}{V_{soil}} .
\label{theta}
\end{equation}
This is also referred to as the soil-water retention curve (SWRC), the degree of saturation, and the volumetric water content.  The characteristic curve allows a direct comparison of the water-holding capacity of soils; see, e.g., Ref.~\cite{Nam10} for example data and empirical fitting forms.  The characteristic curve also contains information about the soil pore structure; see, e.g., Ref.~\cite{Sillers01} for a review of mathematical models.  In spite of this body of work, the experimental accuracy and reproducibility of this approach have been long-standing issues.  Studies \cite{Collisgeorge52, Madsen86, Gee02, Cresswell08, Bittelli09} have shown that the soil water content results obtained from this technique vary when different operating procedures and measuring time scales are used.  Most of these previous studies focus on high gas pressures, under which the equilibrium state become extremely hard to reach and its influence limits accuracy. In this regime flow and dynamic effects due to viscosity, including fingering instabilities, play an important role \cite{LovollPRE04, ToussaintEPL05, LovollTMP11}.  Less attention has been paid to the influence of soil packing height in comparison with a natural capillary rise height, and to operation under low driving pressure essentially at hydrostatic static equilibrium; this is our focus.  Note however that the system is not in thermodynamic equilibrium, since the expulsion of water is not reversible due to hysteresis due to microscopic effects such as contact angle hysteresis and a pressure threshold to move the air-water interface between adjacent pores.

Mercury porosimetry \cite{Dullien1970, Allen, Leon98, Giesche06} is a popular method for characterizing the pore structure for rocks, rather than soils, but has close parallels to the pressure plate method.  For mercury porosimetry measurements, the mercury is forced to penetrate into the pores of a dry sample as the gas pressure is reduced.  The variation of the intruding mercury volume with the reduced gas pressure gives the cumulative pore size distribution of the sample.  Data analysis relies upon two assumptions: first, that the sample pores have a cylindrical geometry; second, that the pressure difference due to the sample packing height can be ignored when compared to the applied pressure \cite{Klinkenberg57, Dong05, Giesche06}.  However, due to the high density of the mercury (e.g., $1$~cm mercury column corresponds to a pressure of $1.3$~kPa.), the second assumption may cause large deviations when data is obtained at low pressure values or from a relatively high sample packing.  Considering the similarity between the pressure plate measurement and the mercury intrusion measurement, we may convert the soil water retention data to the soil pore size distribution in the same way, except that the advancing contact angle of mercury used in the deduction should be replaced by the receding contact angle of water.

In this paper we build a custom pressure plate apparatus for measuring the volume of the expelled water from a soil sample as a function of applied pressure $P$.  Rather than use the characteristic curve $\theta(P)$ to analyze the results, we introduce a new dimensionless parameter, the differential expelled water curve
\begin{equation}
	E(P) = {\rho g \over A_0} { {\rm d}V_w \over {\rm d}P },
\label{expwater}
\end{equation}
where ${\rm d}V_w$ is the incremental volume of water expelled by increasing the pressure across the sample from $P$ to $P+{\rm d}P$, $A_0$ is the sample cross-sectional area, $\rho$ is the density of the expelled water, and $g=9.8$~m/s$^2$.  To analyze the results, we develop our own capillary bundle model for extracting the area distribution of pore radii from the $E(P)$ curves.  The bundles are vertical, rather than horizontal \cite{Klinkenberg57}, and their height plays an important role for water retention that, we emphasize, must be explicitly accounted for in order to correctly analyze pressure plate data.  Failure to do so would introduce a systematic error, and hence an uncontrolled source of irreproducibility, that we make obvious.

The paper is organized as follows.  We begin by describing our custom pressure plate apparatus, the model soils, and the procedures for taking data.  Then we introduce a capillary bundle model that directs the extraction of the water-accessible pore size distribution from experimental data.  The validity of the model is demonstrated by comparison of results obtained for model sandy soil samples with several different packing heights, which have extremely different $E(P)$ curves.  Finally, the samples of uniformly mixed model soils and hydrogel particles are examined.  The effects of gel concentration, gel size, and external confinement on the soil pore structure are determined respectively.


\section{Experiment}

\begin{figure}
\includegraphics[width=2.75in]{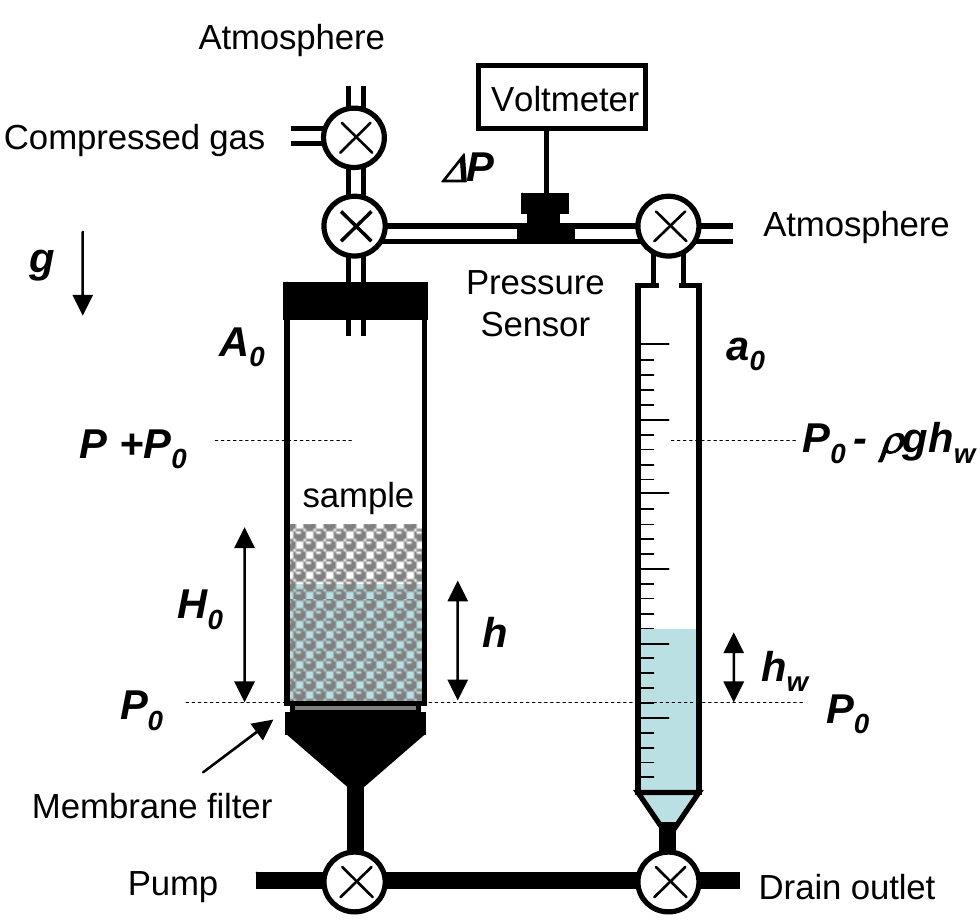}
\caption{(Color online)  Schematic of the custom pressure plate measurement set-up.  A cylindrical glass column with inner cross-sectional area of $A_0$ holds a soil sample packing with a height of $H_0$.  It connects to a pump and a burette from the bottom.  The pump is used to pre-saturate the soil sample, while the burette with an inner cross-sectional area of $a_0$ collects the water expelled out during the pressurization.  A compressed-gas source is used to pressurize the sample column.  The amount of the pressure in sample column that is higher than the gas phase in burette ($\Delta P$) is measured by a differential pressure sensor combined with a high-resolution voltmeter.  The pressure across the sample packing is then determined as $P=\Delta P-\rho gh_w$ (Eq.~(\ref{P})).  For the fixed volume experiments, centimeter-size balls are added on the top of the sample packing to fill the remaining empty space of the sample column before pre-saturation; this prevents expansion of the medium due to swelling of the hydrogel particles when the sample is wetted. }
 \label{Setup}
\end{figure}

Our custom pressure plate apparatus for measuring $E(P)$ curves is illustrated schematically in Fig.~\ref{Setup}.  A cylindrical glass column (Knotes, NJ) holds the soil sample.  It is $30$~cm height with a constant inner cross-sectional area of $A_0=18~$cm$^2$, and is designed to safely pressurize up to $340$~kPa.  Two PTFE end fittings with $20~\mu$m porosity polyethylene bed supports are supplied to seal the top and the bottom of this sample column.  The bottom of the sample column connects to a gear pump (Micropump Inc.), a collection burette (Knotes, NJ), and a drain outlet through two three-way valves and Tygon tubes.  The pump can provide flow rates ranging from $3$ to $60$~mL/min.  It is used to pre-saturate a soil sample by a slow upward infiltration of water from below.  The collection burette is about $40$~cm height with an inner cross-sectional area of $a_0=7.5$~cm$^2$.  It collects the water expelled from the soil sample during the pressurization.  The top of the sample column connects to atmosphere, a compressed-gas source (cylinder of compressed N$_2$, Airgas Inc.), and a differential pressure sensor (26PCA, Omega, CT) through two three-way valves.  The output of the pressure sensor ($\Delta P$) is measured by a voltmeter (Keithley Inc.) with a resolution of $0.1$~mV.  We calibrate the pressure sensor by water columns with controlled heights and get a linear dependence with a sensitivity of $0.29$~kPa/mV.

\subsection{Materials}

The model sandy soil we choose are glass beads (Potters Industries, PA) with a diameter of $2$~mm ($\pm10$\%).  To clean the surface, they were first burnt in a furnace at $500^{\circ}$C for $72$ hours and then soaked in a $1$~M HCl bath for an hour.  After that, the beads were rinsed with deionized (DI) water thoroughly and baked in a vacuum oven at about $110^{\circ}$C for $24$ hours.  The dry glass beads have very hydrophilic surfaces.

The hydrogel particles used in the experiments are a commercial product provided by Degussa Inc.(Stockosorb SW), made by grinding a bulk gel.  The particle shape is randomly faceted but compact.  The main chemical component of these particles is cross-linked polyacrylamide-co-potassium acrylate.  A small amount of salts is present from the industrial polymerization.  If allowed to freely swell in DI water (the salt concentration in the final fluid is less than $10^{-3}$~M) under atmosphere, a gel particle can hold several hundred times of water than its weight in dry.  And $95$\% of the absorbed water is available to plants \cite{Datasheet}.  In our experiments, two different sizes of dry hydrogel particles are chosen.  They come from the same sample bag but were sieved by different sized copper meshes.  The smaller ones ($0.2-0.3$~mm diameter) are used in most of the measurements, while the larger ones ($0.9-1.1$~mm diameter) are only used for comparison.  For comparison, the tetrahedral hole for our glass beads has diameter $0.225\times2$~mm = 0.45~mm.

The gel particles are mixed into the glass beads at four different concentrations: 0.01, 0.05, 0.10, and 0.20 weight percent.  The corresponding gel:bead number ratios are about 1:15, 1:3, 2:3, and 4:3.  To ensure good mixing, a small amount of sample is prepared at a time.  For the desired concentration, carefully weighted dry gel particles are poured into 50 grams of dry glass beads and the entire volume is thoroughly stirred in a large bowl.  The mixture is then gently poured into the sample column.  The process is repeated typically three times, until enough mixture is obtained to fill the sample column to the desired height.  After the sample is wetted, the hydrogel particles become more visible due to the swelling, and we can visually confirm uniform mixing.

Since water is pulled upwards against gravity into a dry hydrophilic sample, other important parameters for our system include the liquid mass density $\rho=1$~g/cm$^3$, the liquid-gas surface tension $\gamma=73$~dyne/cm, and the contact angle $\theta=0^\circ$ between the liquid-gas interface and the hydrophilic grain surfaces.

\subsection{Procedures and example data}

In this subsection we describe how our samples are prepared and how our apparatus is used to measure both the pressure $P$ and the dimensionless differential expelled water curve $E(P)$ vs $P$.

To prepare a soil packing, we first clean and dry the sample column to make sure that the inner surface is hydrophilic.  A piece of filter paper (Whatman, NJ) is added on the bottom and the dry granular sample is poured into the sample column carefully, $1$-$2$~cm height each time, until reaching the desired packing height $H_0$.  The sample column is then lightly patted so that the top surface is level and the packing fraction of grains is $0.62\pm0.01$, within error of random-close packing \cite{DuranBook}.  Two different conditions are used for each mixture:  fixed volume and free swelling.  For fixed volume, centimeter-size plastic balls are added on the top of the granular packing to fill the remaining empty space in the sample column and maintain the packing at constant volume against the pressure of the swelling hydrogel particles.  For free swelling, no plastic balls are added and the sample column has enough empty space for the mixture packing to freely expand when the intially-dry hydrogel particles absorb water.

The packing is then pre-saturated with DI water from the bottom at a slow flow rate of $3$~mL/min with the top open to atmosphere.  For the pure glass beads packings, the pre-saturation procedure does not modify the pore structure and the extra confinement has no effect on the results.  However, for the mixed packings of hydrogel particles and glass beads, the swelling hydrogel particles tend to expand the packing during the pre-saturation procedure, so free versus fixed volume conditions are different.   After pre-saturation, each mixture packing is left for 24~hours to ensure the full swelling of the hydrogel particles.

The final preparation step is to open the top of the sample and collection columns to atmosphere, and to drain liquid in the collection column down to the same level as membrane filter at the bottom of sample column -- i.e. the location indicated by a horizontal line labeled $P_0$ in Fig.~\ref{Setup}.  Once this level is reached, the drain outlet is closed; however, over the course of several hours, the liquid level in the sample column falls and the level in the collection column rises.  This extra liquid is then drained, and the whole process is iterated as many times as necessary so that the liquid level in the collection column remains constant and even with the bottom of the sample column.  When this is finally achieved, both columns are sealed at top.  Referring to the quantities labeled in Fig.~\ref{Setup}, the initial conditions are thus such that $P_0$ equals atmospheric pressure, $P=0$, $\Delta P=0$, $h_w=0$, and $h$ equals the equilibrium capillary rise of water pulled into the sample against gravity.

The dimensionless expelled water curve, $E(P)$ vs $P$, is now measured by pressurizing the space above the sample and measuring the resulting expulsion of water from the increase in the height $h_w$ of water in the collection burette.  This is done in a step-wise fashion, by repeatedly bleeding in a small quantity of compressed gas (N$_2$) and then waiting for the liquid levels to come to equilibrium before $h_w$ is recorded.  The two quantities directly measured are thus the pressured difference $\Delta P$ between the columns, and the height $h_w$ of liquid in the collection burette.  In order to deduce the pressure $P$ across the sample, first note that the pressure in the collection burette at the level of the membrane filter is the same value, which we call $P_0$, as at the bottom of the sample.  This reference pressure is now greater than atmospheric, but its value is not of interest.  The gas pressure in sample column equals $P_0+P$, and the gas pressure in the collection burette equals $P_0-\rho g h_w$ (see Fig.~\ref{Setup}).  The gas pressure in the sample column is also greater than that in the collection burette by the measured quantity $\Delta P$.  Altogether this gives the pressure across the sample as
\begin{equation}
   P = \Delta P - \rho gh_w.
\label{P}
\end{equation}
Note that $P$ gives the amount by which the pressure is greater in the gas above, than in the liquid underneath, the sample -- and hence is sometimes referred to as ``suction".  In order to deduce $E(P)$, note that the incremental volume ${\rm d}V_m$ of water expelled by a small increase in $P$ is simply the product $a_0{\rm d}h_w$ of the inner area of the collection burette times the change in collected water level.  The dimensionless differential expelled water curve for our apparatus is thus
\begin{equation}
	E(P) = (a_0/A_0)\rho g~{\rm d}h_w/{\rm d}P,
\label{ep}
\end{equation}
which may be found by numerical differentiation of $h_w$ versus $P$ data.

\begin{figure}
\includegraphics[width=2.75in]{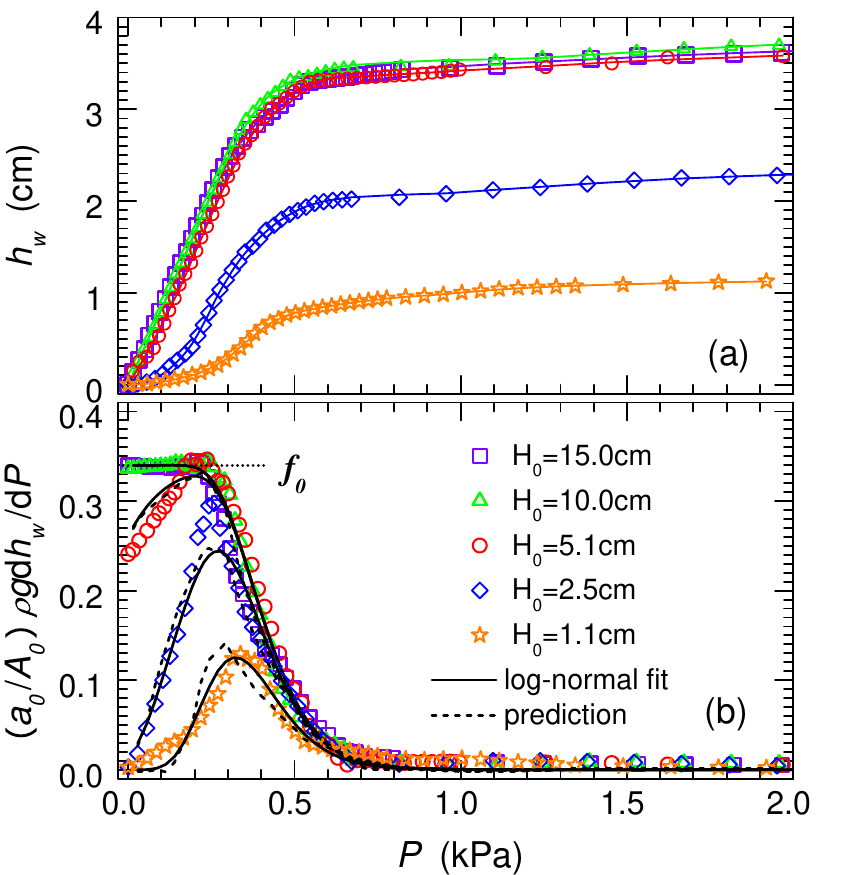}
\caption{(Color online)  (a) Height of water in the collection burette, and (b) dimensionless differential water expulsion parameter, versus applied pressure for 2~mm diameter glass beads packed to different heights $H_0$ as labeled.   As shown in Fig.~\ref{Setup}, $A_0$ and $a_0$ are the inner cross-sectional areas of the sample column and the burette, respectively  The y-axis in (b) represents the expelled water per pressure increment per sample area, made dimensionless by the fluid density $\rho$ and $g=9.8$~m/s$^2$.   In (a) the solid curves represent the smoothed data used for differentiation.  In (b) the solid curves are a simultaneous fit to Eq.~(\ref{unitWater2}) for a log-normal pore size distribution with $f(r)$ given by Eq.~(\ref{Lognormal}), and the dashed curves are the expectation for the low packing height data based on the high packing height data and Eq.~(\ref{unitWater2}).  The plateau in (b) labeled $f_0$ represents the cross-sectional area fraction of water-accessible pore space in the packing.}
\label{PackingHeight}
\end{figure}

Example data for $h_w$ versus $P$ obtained by the above procedures are shown in Fig.~\ref{PackingHeight}a, for 2~mm glass beads (no hydrogel particles) packed to different heights $H_0$ as indicated in the legend.  For all, $h_w$ increases monotonically with $P$ towards an asymptotic value corresponding to the complete expulsion of {\it all} water from the sample.  To reach this limit, the presence of a wet filter paper membrane beneath the sample was necessary to prevent the penetration of compressed gas out from underneath the sample.  Note that the results for the highest packing heights are nearly identical, and display an initial rise from zero that is linear.   But the results for the lowest packing heights are sigmoidal in shape, and asymptote to values that decrease for smaller $H_0$.  Thus the soil-water characteristic curves (SWCC), given by Eq.~(\ref{theta}) as $\theta(P)=1-h_w(P)/h_w(\infty)$, clearly depend on packing height.  This can be understood as follows.  When the packing is taller than the equilibrium capillary rise,  the upper portion of the sample is dry and has no influence on water retention.  When the packing is smaller than the equilibrium capillary rise, the entire sample is wet and a minimum height-dependent pressure head must be exceeded in order for water expulsion to commence.  This intuition will be made quantitative with a capillary bundle model in the next section.

The differential dimensionless expelled water curves, $E(P)$ versus $P$, obtained from the example $h_w$ versus $P$ data and Eq.~(\ref{ep}), are plotted in Fig.~\ref{PackingHeight}b.  For numerical differentiation, we first smooth the data using the LOESS algorithm~\cite{Cleveland79} available in Igor.  This fits the data to a quadratic polynomial by sub-regions of size set by a user-specified smoothing parameter.  The advantage of this algorithm is that it is very flexible and does not require a specific function to fit the entire data set.  During the smoothing process, we vary the smoothing parameter between 0.25 to 0.5 to obtain the smoothest curve that does not systematically deviate from the data.  Then we do central finite differencing on both the smoothed curve and the original data  to ensure that the smoothed curve describes the original data well even after differentiation.  For small packing heights, the $E(P)$ curves rise to a peak and then fall toward zero as $P$ increases.  For the two tallest packings, the $E(P)$ results are indistinguishable, monotonically decreasing with $P$, are equal to a constant $f_0\approx0.34$ for small $P$.  The value of $f_0$ represents the total cross-sectional area fraction of the water accessible pore space, as will be shown next with a capillary bundle model.


\section{Capillary bundle model}

\begin{figure}
\includegraphics[width=2.75in]{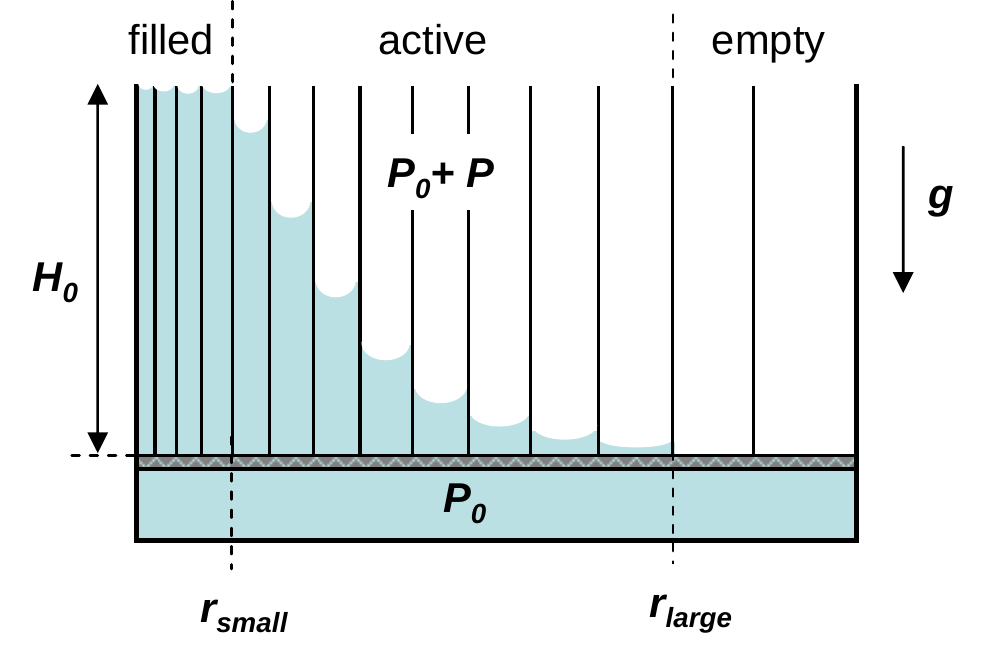}
\caption{(Color online) Schematic of the capillary bundle model.  In the model, a bundle of capillaries with height of $H_0$ is placed vertically in a water reservoir.  The pressure in the water reservoir is $P_0$ and the pressure in gas phase is $P_0+P$ ($P\geq0$).  Here, $r_{small} = 2\gamma /(P+\rho gH_0)$ sets the radius limit for the filled capillaries; $r_{large} = 2\gamma /P$ sets the radius limit for the empty capillaries; and the capillaries with radius between them are the active ones.  Only the water in the active capillaries is expelled out when an small pressure increment d$P$ is applied to the system. }
\label{Model}
\end{figure}

To extract physical meaning from the differential $E(P)$ versus $P$ water expulsion curves, we now construct a model in which the granular packing is pictured as a set of vertical capillary tubes of height $H_0$ and with some distribution of radii $r$, as depicted schematically in Fig.~\ref{Model}.  Such ``capillary bundle'' approximations may seem rather severe and uncontrolled, but they have a long history of use in the modeling of fluids in porous media \cite{Klinkenberg57, Rao76, Chu93, Dahle05, Dong05, Choi08}, including evaporative drying \cite{LehmannPRE08}.   Here, the rise $h$ of liquid into a tube of radius $r$ may be computed by considering how the pressure increases from $P_0$ in the liquid at the bottom of the sample to $P_0+P$ in the gas above.  In going upwards from the bottom to a height $h$ just below the liquid-gas interface, the pressure drops according to Pascal's law by $\rho g h$.  And in crossing the interface, the pressure goes up according to Laplace's law by $2\gamma/r$ where $\gamma$ is the liquid-gas surface tension and where complete wetting is assumed.   In other words, the gas pressure $P_0+P$ above the sample is equal to $P_0-\rho g h + 2\gamma/r$, and this gives the capillary rise as
\begin{equation}
	h = {2 \gamma \over \rho g r} - {P \over \rho g}.
\label{hi}
\end{equation}
The first term represents the usual capillary rise formula, which would be multiplied by the cosine of the contact angle for the case of partial wetting.  The second term represents the reduction in height due to an applied pressure (or suction), and is independent of $r$ and wetting properties.

Note that Eq.~(\ref{hi}) holds only if $r$ is neither too small nor too large.  In particular if $r$ is smaller than
\begin{equation}
   r_{small} = \frac{2\gamma}{P+\rho gH_0}
\label{rs}
\end{equation}
then the tube is too short and the rise of fluid will be pegged at $h=H_0$, with water entirely filling the tube.  And if $r$ is larger than
\begin{equation}
  r_{large} = \frac{2\gamma}{P},
\label{rl}
\end{equation}
then the applied pressure is too great and the rise will be pegged at $h=0$, with all water completely expelled from the tube.  Tubes of radii in the range $r_{small} < r < r_{large}$ are partially filled with fluid, and are ``active'' in the sense that their filling height $h$ responds to pressure changes according to Eq.~(\ref{hi}).

The simplest case to consider, first, is a sample with tall packing height and with small applied pressure -- then all tubes are active.  Therefore, in response to a small increase ${\rm d}P$ of applied pressure, the change in rise height for {\it all} tubes is given by differentiating Eq.~(\ref{hi}) as ${\rm d}h = -{\rm d}P/(\rho g)$.  The resulting volume of expelled water is $-f_0 A_0 {\rm d}h$ where $A_0$ is the cross-sectional area of the sample and $f_0$ is the total cross-sectional area fraction of the pore space.  By continuity, the change ${\rm d}h_w$ in height of liquid in the collection burette is such that the expelled volume equals  $a_0{\rm d}h_w$.  In other words, we have $a_0{\rm d}h_w = -f_0 A_0 [-{\rm d}P/(\rho g)]$; therefore, when all tubes are active, the cross-sectional area fraction is
\begin{equation}
	f_0 = (a_0/A_0)\rho g~{\rm d}h_w/{\rm d}P,
\label{f0}
\end{equation}
which is recognized as our dimensionless differential water expulsion parameter.  This holds for tall samples and low pressures, for which the $E(P)$ versus $P$ sample data in Fig.~\ref{PackingHeight}b are indeed constant.   For that sample, the inferred total cross-sectional area fraction of the pore space may be read off the graph as $f_0=0.34\pm0.01$.  Note that this argument relies only upon continuity and the second term in Eq.~(\ref{hi}); therefore, we believe its validity transcends any limitations of the capillary bundle approximation.

Now we generalize to the case that only some of the tubes are active.  For this we introduce a new quantity, {\it the fraction $f(r)$ of the cross-sectional area having pores with radii less than $r$}.  By definition, $f(r)$ increases monotonically from 0 and asymptotes to $f_0$ as $r$ increases from zero to infinity.  Also by definition,  $f(r)$ is a cumulative distribution function and therefore the associated probability distribution function (PDF) is
\begin{equation}
   p(r) = \frac{{\rm d}f}{{\rm d}r},
\label{pdf}
\end{equation}
which is normalized to $f_0$ rather than to one.  As above, the volume of water expelled by an increase ${\rm d}P$ of applied pressure is $a_0 {\rm d}h_w$ and equals the active area times the decrease in liquid level inside the sample, $-{\rm d}h = {\rm d}P/(\rho g)$.  Whereas before the active area was $f_0 A_0$, it is now more generally $[f(r_{large})-f(r_{small})]A_0$ where the term in square brackets is the fraction of active area with radii in the range $r_{small} < r < r_{large}$ that obey Eq.~(\ref{hi}).  Altogether the capillary bundle model thus gives
\begin{eqnarray}
	(a_0/A_0)\rho g{ {\rm d}h_w \over {\rm d}P }
   		&=& f(r_{large})-f(r_{small}), \label{unitWater}  \\
     		&=& f\left(\frac{2\gamma}{P}\right)-f\left(\frac{2\gamma}{P+\rho gH_0}\right), \label{unitWater2}  \\
		&=& \int_{2\gamma /( P+ \rho g H_0)}^{2\gamma/P} p(r){\rm d}r .\label{unitWater3}
\end{eqnarray}
where the left hand side is recognized as our dimensionless differential water expulsion parameter, $E(P)$, and the large and small radii are taken from Eqs.~(\ref{rs}-\ref{rl}).  Note that the right hand sides reduce to $f_0$, and Eq.~(\ref{f0}) is recovered, in the limit that all pores are active such that the integration limits lie between $r_{small}$ and $r_{large}$.  These equivalent expressions are the main result of the capillary bundle model.  In essence, the raw data from pressure plate measurements of $h_w$ versus $P$ are seen to be directly linked to a double integral of the cross-sectional area distribution of the pore radii.

As a remark, note that while our model thus shows that the {\it cross-sectional area} distribution of pore radii is the key structural quantity accessible from water retention/expulsion data, prior work has been in terms of the {\it volumetric} distribution of pore radii~\cite{Sillers01}.


\section{Model Sandy Soil}

In this section we describe how to use the capillary bundle model, Eqs.~(\ref{unitWater}-\ref{unitWater3}), to deduce pore size information from experimental data for $E(P)$ versus $P$.  And we demonstrate the procedures using the example data of Fig.~\ref{PackingHeight} for packings of 2~mm glass spheres of various heights.

\subsection{Direct fitting}

One straightforward method of analysis is to assume a particular form for $f(r)$ and then simply fit $E(P)=(a_0/A_0)\rho g{\rm d}h_w/{\rm d}P$ data to the right hand side of Eq.~(\ref{unitWater2}).  In soil science, the approach is often to perform a similar empirical fit to the soil water retention curve.  By contrast, we work not with the measured quantity but with an underlying quantity of direct physical significance.  In absence of theoretical guidance, we try three different empirical forms of $f¨$ for which the PDF is a peaked function:
\begin{eqnarray}
   f(r) &=& f_0[1-e^{-(r/r_0)^\alpha}] , \label{Sigmoid1} \\
   f(r) &=& f_0[1-(1+(r/r_0)^\beta)e^{-(r/r_0)^\beta}] , \label{Sigmoid2} \\
  f(r) &=& \frac{f_0}{\sqrt{2\pi{\sigma_0}^2}}\int_{0}^{r}x^{-1}e^{-\frac{(\ln x-\ln r_0)^2}{2{\sigma_0}^2}}\, {\rm d}x . \label{Lognormal}
\end{eqnarray}
All three cumulative distribution functions are sigmoidal in shape, rise from $f(0)=0$, and asymptote to $f_0$ as $r\rightarrow\infty$.  The last form, Eq.~(\ref{Lognormal}), is the cumulative distribution function for a log-normal PDF.  These forms are used to directly, and simultaneously, fit the data for all five packing heights in Fig.~\ref{PackingHeight}b by keeping $f_0=0.34$ fixed and adjusting the other two parameters.  The fits are all satisfactory and give $r_0=0.40$~mm and $\alpha=4.2$ for Eq.~(\ref{Sigmoid1});  $r_0=0.30$~mm and $\beta=2.7$ for Eq.~(\ref{Sigmoid2}); and $r_0=0.36$~mm and $\sigma_0=0.29$ for Eq.~(\ref{Lognormal}).   The log-normal fits are displayed in Fig.~\ref{PackingHeight}b and have the smallest chi-squared deviation of the three candidate forms.  A log-normal form is consistent with simulation results for the pore size distribution for a random packing of spheres \cite{Yang06, Yu07, Reboul08}.    We emphasize that, while the $E(P)$ versus $P$ curves are all very different and height dependent, in effect they all give the same pore radii distribution.  The good agreement of the simultaneous fits for {\it all} data sets demonstrated both the consistency of our data and validity of the capillary bundle model.

\subsection{Extraction of $f(r)$ for tall packings}

\begin{figure}
\includegraphics[width=2.75in]{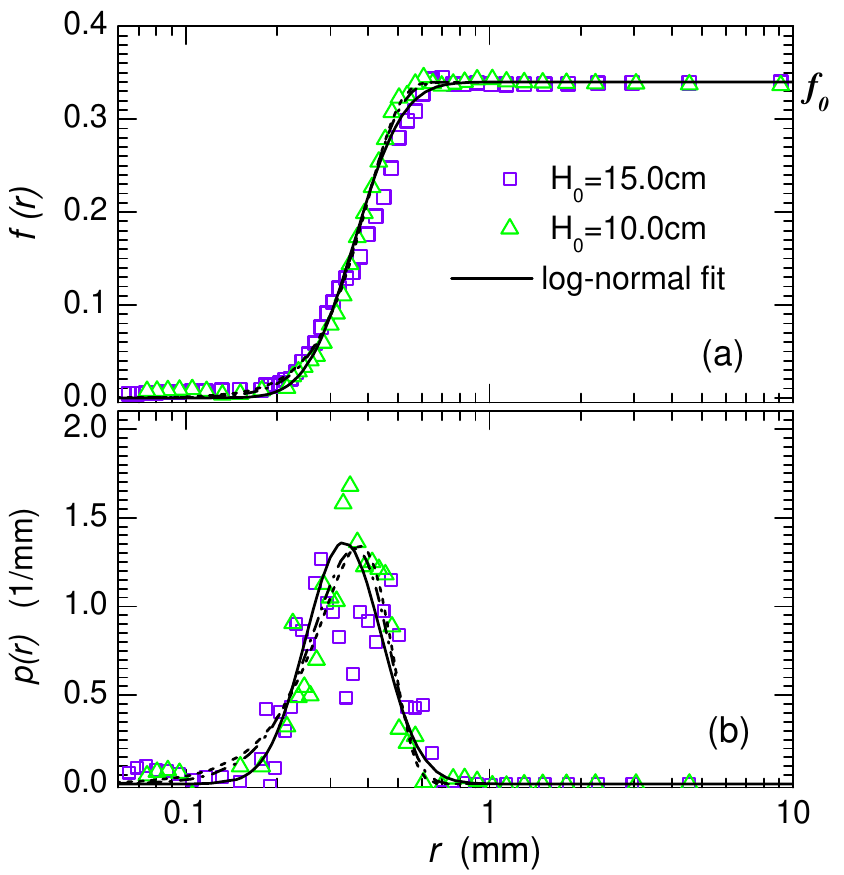}
\caption{(Color online)  Variation of (a) cumulative area fraction of water-accessible pore space with pore radii smaller than $r$ and (b) associated probability density, which is normalized to the total area fraction $f_0$ of the pore space.  These results are based on the water expulsion data of Fig.~\ref{PackingHeight} for $2$~mm glass beads with packing heights $H_0$ as labeled.  These samples are tall enough that $f(r)$ could be extracted directly by use of Eq.~(\ref{frHP}).  The solid curve is the log-normal form, Eq.~(\ref{Lognormal}), found by simultaneous fit of Eq.~(\ref{unitWater2}) to all data in Fig.~\ref{PackingHeight}.  The dashed curves are similarly obtained fits to Eqs.~(\ref{Sigmoid1}-\ref{Sigmoid2}).  The PDF equals ${\rm d}f/{\rm d}r$, and the data points were computed by finite differencing with no further smoothing.}
\label{HighPacking}
\end{figure}

For large enough packing heights $H_0$, such that the capillary rise of liquid never extends to the top of the sample, $f(r_{small})$ vanishes and the right hand side of Eq.~(\ref{unitWater2}) reduces to $f(2\gamma/P)$.  Then the cumulative area fraction of pore space with radii less than $r$ is given by
\begin{equation}
	f(r) \big\vert_{2\gamma/P}  =  (a_0/A_0)\rho g~{\rm d}h_w /{\rm d}P,
\label{frHP}
\end{equation}
and a plot of $f(r)$ versus $r$ is obtained directly by plotting $E(P)=(a_0/A_0)\rho g{\rm d}h_w/{\rm d}P$ data versus $2\gamma/P$.  Results are shown in Fig.~\ref{HighPacking}a based on the Fig.~\ref{PackingHeight}b water expulsion curves for the two tallest packings.  For comparison, the direct fitting results for the three candidate sigmoidal forms are also included.  The associated probability distributions obtained by differentiation, for both the data and the fits, are shown in Fig.~\ref{HighPacking}b.

\subsection{Extraction of $p(r)$ for short packings}

\begin{figure}
\includegraphics[width=2.75in]{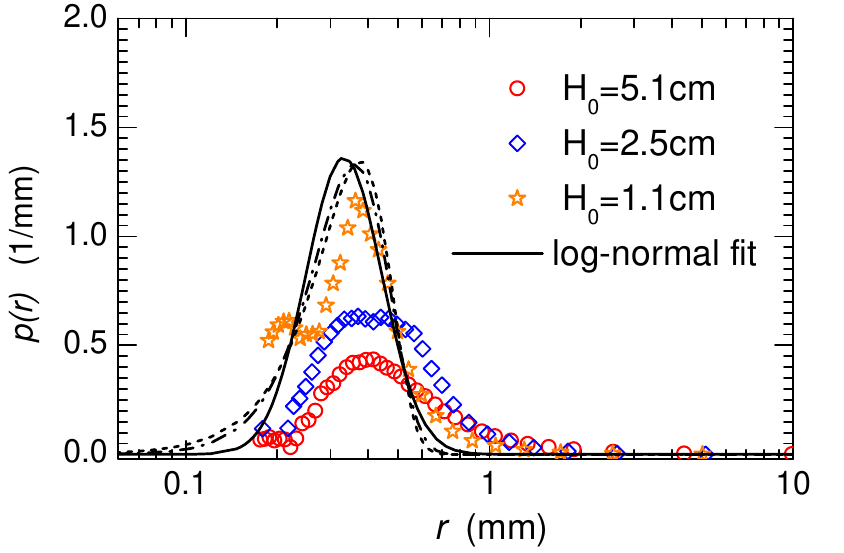}
\caption{(Color online) The distribution of the area fraction of pore space for $2$~mm glass beads, deduced from Eq.~(\ref{shortPDF}) using the water expulsion data of Fig.~\ref{PackingHeight} at the three shortest packing heights $H_0$ as labeled. The log-normal fit was obtained previously, by the simultaneous fit of all data in Fig.~\ref{PackingHeight}.[Note: the data cut-off used in this figure is set to be $P=0.8$ kPa]}
\label{ShortPacking}
\end{figure}

For small enough packing heights, such that $\rho g H_0 \ll P$, the right hand side of Eq.~(\ref{unitWater}) may be well approximated as ${\rm d}f/{\rm d}r=p(r)$ evaluated at $r=2\gamma/P$ and multiplied by $r_{large}-r_{small}\approx 2\gamma \rho g H_0/P^2$.  Thus the radius distribution for the area fraction of pore space is given as
\begin{equation}
	p(r)\big\vert_{2\gamma/P}  = {  (a_0/A_0)\rho g~{\rm d}h_w/{\rm d}P   \over  (2\gamma/P)(\rho g H_0/P) }.
\label{shortPDF}
\end{equation}
In other words, a plot of $p(r)$ versus $r$ is obtained directly by dividing the $E(P)$ data by the length $(2\gamma/P)(\rho g H_0/P)$ and plotting versus $2\gamma/P$.  The right hand side of Eq.~(\ref{shortPDF}) is computed for the three shortest packing height data of Fig.~\ref{PackingHeight}, for the $2$~mm glass spheres, and is plotted in Fig.~\ref{ShortPacking} along with the log-normal distribution found from previous fits.  We see that results become spurious at small $r$.  But more importantly, for larger $r$, we see that the results underestimate the expectation but become progressively better for the smaller packing heights.  Therefore, analysis of water expulsion curves with Eq.~(\ref{shortPDF}) would require even shorter samples than measured here.

\section{Soft Hydrogel Particles As Soil Additives}

\subsection{Gel concentration}

\begin{figure}
\includegraphics[width=2.75in]{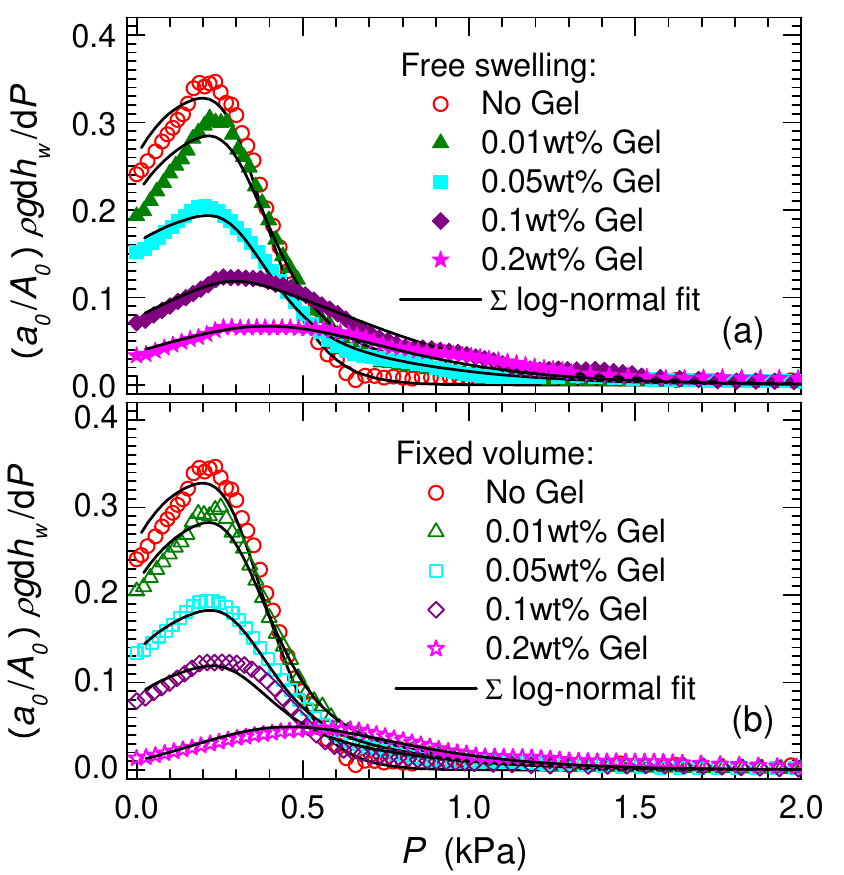}
\caption{(Color online)  Dimensionless differential water expulsion parameter, versus applied pressure for $2$~mm diameter glass beads with different concentration small hydrogel particles in (a) free swelling condition and in (b) fixed volume condition.  The solid curves are a simultaneous fit to Eq.~(\ref{unitWater2}) for a combined log-normal pore size distribution with $f(r)$ given by Eq.~(\ref{Lognormal2}). The fitting parameters are shown in Fig.~\ref{PDFgel} and Fig.~\ref{FittingParameters}. }
\label{WaterExpulsion}
\end{figure}

When hydrogel particles are uniformly mixed into sandy soils, the pore structure is modified according to both the concentration and the size of additives.  Small gel particles ($0.2-0.3$~mm axis in dry) are used to examine the influence of gel concentration.  In dry, they can fit into the existing soil pores without disturbing the soil matrix, since the tetrahedral hole for 2~mm diameter beads is about 0.45~mm.  After pre-saturation of the sample, however, their maximum swelling size may exceed the pore size.  Whether they can reach this size or not depends on the strength of the soil matrix confinement during pre-saturation.  Fig.~\ref{WaterExpulsion} shows the $E(P)$ data at four different gel concentrations in both (a) free swelling and (b) fixed volume conditions. The data for a ``no gel'' packing is included for comparison.  As the gel concentration increases, it become harder and harder to expel water out of the soil packing.  Note that there is no dramatic difference between free swelling data and fixed volume data.

Direct fitting is applied to extract the water-accessible pore structure in these mixture packings.  Data in Fig.~\ref{WaterExpulsion}(a) and in Fig.~\ref{WaterExpulsion}(b) are fitted simultaneously by following function respectively:
\begin{widetext}
\begin{eqnarray}
  f(r) = f_0\left(\frac{\delta}{\sigma_0\sqrt{2\pi}}\int_{0}^{r}x^{-1}e^{-\frac{(lnx-lnr_0)^2}{2{\sigma_0}^2}}\, {\rm d}x + \frac{1-\delta}{\sigma_1\sqrt{2\pi}}\int_{0}^{r}x^{-1}e^{-\frac{(lnx-lnr_1)^2}{2{\sigma_1}^2}}\, {\rm d}x\right).
\label{Lognormal2}
\end{eqnarray}
\end{widetext}
This is a combination of the cumulative distribution functions for two log-normal PDFs.  The first one represents the pores existing between soil particles; while the second one represents the pores that may exist between gel particles or between gel particle and soil particle. In each fit, we fixed $r_0=0.36$~mm and $\sigma_0=0.29$ from the previous section on pure glass bead packings, but let the values of $r_1$ and $\sigma_1$ vary simultaneously for all four set of mixture packing data.  The parameter $\delta$ represents the percentage of the unoccupied soil pores in a mixture packing thus its value depends on gel concentration.  It is allowed to adjust freely for each set of data.  $f_0$ is the maximum value of $f(r)$ at $r\rightarrow\infty$ and represents the total area fraction of water-accessible pore space in a packing.  It also is allowed to adjust freely in fits to each set of data.

\begin{figure}
\includegraphics[width=2.75in]{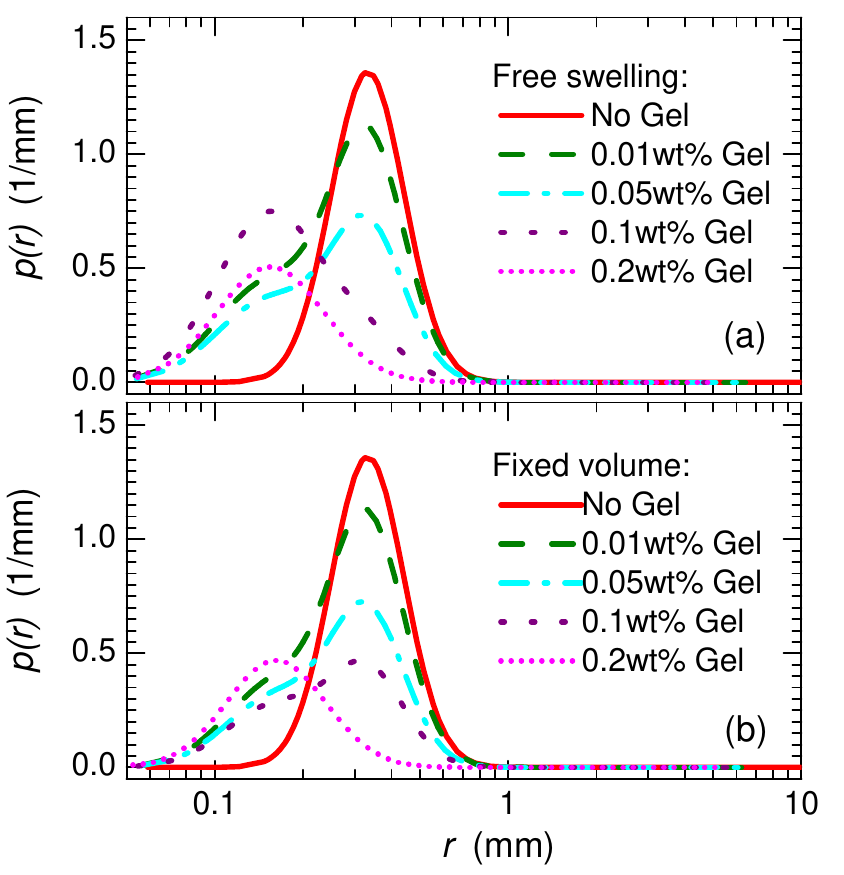}
\caption{(Color online) The distribution of the area fraction of pore space for $2$~mm glass beads with different concentration small hydrogel particles in (a) free swelling condition and in (b) fixed volume condition, deduced from the simultaneous fit of all data in Fig.~\ref{WaterExpulsion}.  In both conditions, the major peak, whose location and width are fixed to be the result in the model sandy soil alone ($r_0=0.36$~mm and $\sigma_0=0.29$), reduces its height as gel concentration increases; while a secondary peak grows in small $r$ region at the same time.  In free swelling condition, the secondary peak is located at $r_1 =0.18$~mm with $\sigma_1=0.42$; in fixed volume condition, it is at $r_1 =0.19$~mm with $\sigma_1=0.39$. }
\label{PDFgel}
\end{figure}

The fitting curves are shown in Fig.~\ref{WaterExpulsion} by solid curves.  We determine that the value of $r_1$ is $0.18$~mm in free swelling condition and $0.19$~mm in fixed volume condition.  The value of $\sigma_1$ is $0.42$ and $0.39$ for free swelling condition and fixed volume condition respectively.  The results obtained in different conditions are very close to each other.  The pores existing between gel particles or between gel particle and soil particle are only about half of the size of the major pores in soil matrix.  For further discussion, we plot the PDFs for packing with different gel concentrations at different conditions in Fig.~\ref{PDFgel}(a) and (b) respectively.  We clearly see that as the gel concentration increases the height of the major peak decreases monotonously and finally disappears when gel concentration exceeds $0.1$~wt\%.  This corresponds to a gel to bead number ratio of about 2:3.  We also notice that the height change of the secondary peak does not follow the same trend as the major one.

\begin{figure}
\includegraphics[width=2.75in]{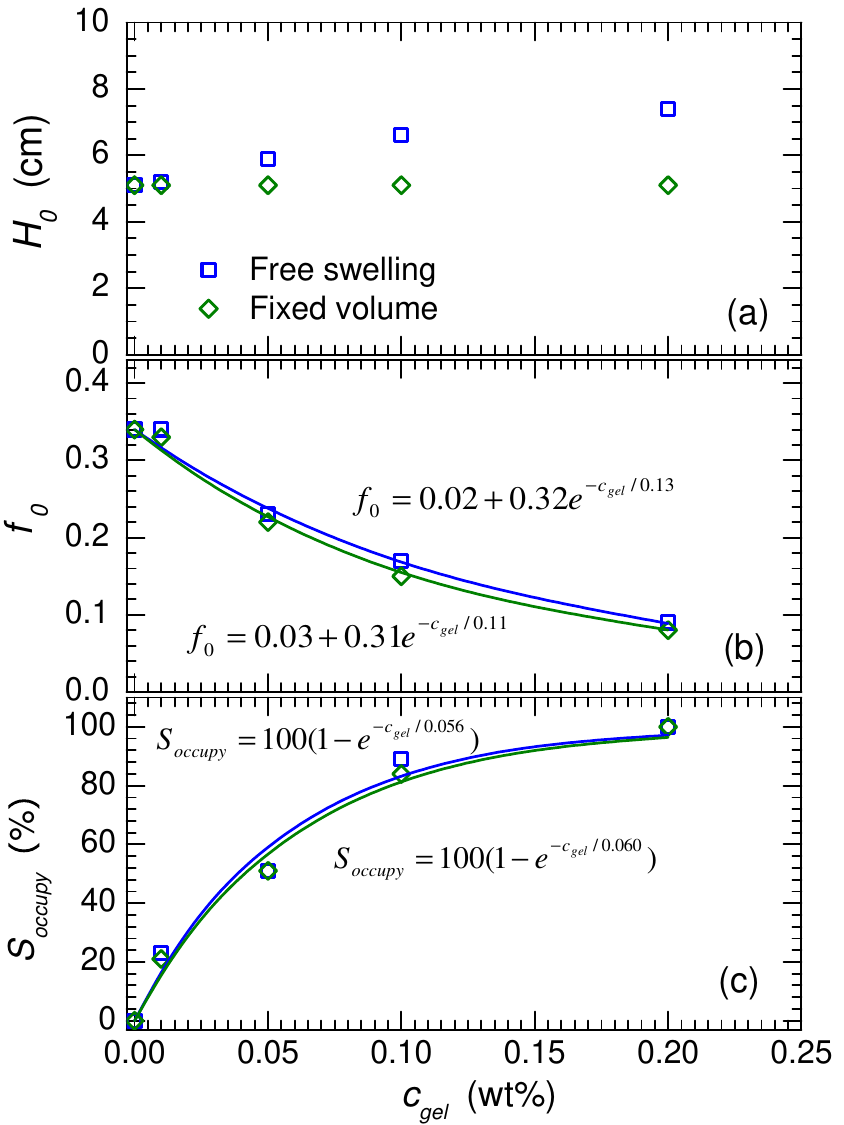}
\caption{(Color online) (a) Packing height and parameters determined from the simultaneous fit shown in Fig.~\ref{WaterExpulsion}: (b) the total area fraction, and (c) effective soil pore occupation rate by gel particles. The error bars are smaller than the symbol sizes.  As gel concentration increases, the total area fraction of the pore space decays exponentially for both free swelling condition and fixed volume condition.  The soil pore occupation rate is defined by Eq.~(\ref{Soccupy}).  It grows almost linearly as gel concentration increases until all major pores are fully-occupied, and hence can be fit to an exponential.}
\label{FittingParameters}
\end{figure}

The behavior as a function of gel concentration is summarized in Fig.~\ref{FittingParameters}.  Fig.~\ref{FittingParameters}(a) shows the packing height, $H_0$, the results of which are used in the fits.  In dry, all the mixture packings have the same height.  When pre-saturated in free swelling condition, the height of the mixture packing varies as the gel concentration varies by up to about thirty percent.  When pre-saturated in fixed volume conditions, all concentration packings are of course forced to maintain their original height.  

Fig.~\ref{FittingParameters}(b) shows the variation of the total area fraction of pore space with gel concentration.  In both conditions, the total area fraction $f_0$ decays exponentially as gel concentration increases.  This result is consistent with our prior studies on the water permeability in the same mixture system \cite{Verneuil11}.  Again, no significant difference is seen between two conditions when correct packing heights are applied.  The characteristic value of gel concentration in the exponential fits is around $0.1$wt\%, which corresponds to the loss of the primary peak in the PDF seen in Fig.~\ref{PDFgel}.

Fig.~\ref{FittingParameters}(c) shows the variation of $S_{occupy}$, the effective percentage of soil pores blocked by gel particles.  This is defined as
\begin{equation}
	S_{occupy} = 100 \left(1-\frac{f_0 \delta}{0.34} \right),
\label{Soccupy}
\end{equation}  
where $0.34$ is the value of $f_0$ at $c_{gel}=0$.  From the figure, we see that the value of $S_{occupy}$ grows linearly, until all soil pores are filled, and thus can be fit to an exponential function of bead concentration.  The gel concentration required to fill all soil pores is around $0.1$wt\%, consistent with the characteristic value obtained from the exponential fit in Fig.~\ref{FittingParameters}(b).  When the gel concentration exceeds this value in free swelling conditions, the swollen gel particles may occupy extra space by expanding the packing.  But in fixed volume conditions, the swollen gel particles have to share that soil pores with others and cannot swell to their desired size.

\subsection{Gel size}

Lastly, to probe the influence of gel particle size, we fixed the gel concentration to be $0.1$~wt\% and compare the results for large gel particle additives ($0.9-1.1$~mm axis in dry) to those above for the small gel particle additives ($0.2-0.3$~mm axis in dry) in the same external conditions.  The large gel particles have a dry size larger than the average soil pore size, and thus may perturb the soil matrix structure even in dry.  After pre-saturation, the free swelling packing with large gel particles expands to a height of $H_0=7.4$~cm while the fixed volume packing maintains a height of $H_0=5.1$~cm.

\begin{figure}
\includegraphics[width=2.75in]{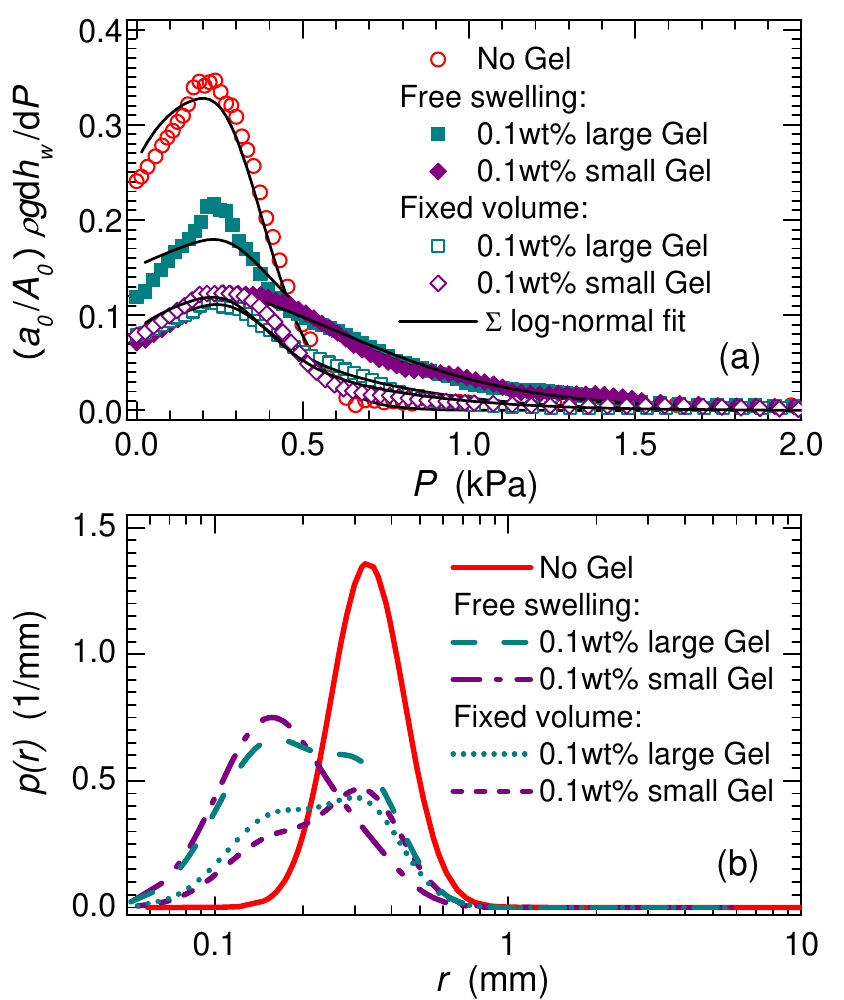}
\caption{(Color online)  (a) Dimensionless differential water expulsion parameter versus applied pressure, and (b) the distribution of the area fraction of pore space for $2$~mm glass beads with $0.1$~wt\% large and small hydrogel particles in both free swelling condition and fixed volume condition.  The solid curves are a fit to Eq.~(\ref{unitWater2}) for a combined log-normal pore size distribution with $f(r)$ given by Eq.~(\ref{Lognormal2}).  The fitting parameters for packings with small gel particle additives have been shown in Fig.~\ref{FittingParameters}.  For packings with large gel particle additives, the values of $r_0$, $\sigma_0$, $r_1$, and $\sigma_1$ are fixed to be the same as those obtained from corresponding small gel packing.  In free swelling condition, the packing height is measured as $H_0=7.4$~cm, the total area fraction is determined as $f_0=0.22$, and the soil pore occupation rate is determined as $S_{occupy}=69$~\%.  In fixed volume condition, the packing height is measured as $H_0=5.1$~cm, the total area fraction is determined $f_0=0.15$, and the soil pore occupation rate is $S_{occupy}=74$\%.   }
\label{SizeEffect}
\end{figure}

Fig.~\ref{SizeEffect}(a) shows the $E(P)$ data for packings with different size gel particles in free swelling and fixed volume conditions, along with the ``no gel'' data for comparison.  Under fixed volume, note that the curves are very similar for the two different gel particle sizes.  Thus, the particle size is relatively unimportant, presumably as long as it is not very much greater than the bead size.  Instead, the weight percent of the additives is the more important parameter for affecting behavior.

Fits to the bimodal log-normal pore size distribution of Eq.~(\ref{Lognormal2}) are also included in Fig.~\ref{SizeEffect}(a).  Since there is only one set of data in each condition, we fix the values of $r_0$, $\sigma_0$, $r_1$, and $\sigma_1$ to be the same as those obtained from the fits of small gel packings.  In free swelling condition, the total area fraction is determined as $f_0=0.22$ and the soil pore occupation rate is determined as $S_{occupy}=69$\%.  Comparing to the corresponding small gel packing, the large gel one has a higher total pore area fraction of space and a lower soil pore occupation rate.  In fixed volume condition, we determine that the total area fraction is $f_0=0.15$, which is very close to that of a small gel packing.  The soil pore occupation rate is determined as $S_{occupy}=74$\%, which is also lower than that in the small gel packing.

The resulting fitting function for the PDFs are plotted in Fig.~\ref{SizeEffect}(b).  From this figure, we clearly see that under free swelling condition more soil pores are left unoccupied in the packing when its small gel particle additives are replaced by the same mass of large gel particles.  However, this is a relatively small effect.  By contrast, the role of external confinement is more important.  Namely, a strong external confinement helps the gel particles to swell into and efficiently block the pore spaces.


\section{Conclusion}

In summary, we developed a custom pressure plate apparatus for measuring the expulsion of water from model soils as a function of applied pressure, and we developed a capillary bundle model to analyze the results in terms of pore-scale structure.  We verified the apparatus and the model by obtaining consistent results for the pore structure for packings of different heights, where the soil-water characteristic curves are all different.  And we applied our methods to study the effect of superabsorbent gel particle additives, which can swell to block the pores.

One general conclusion is that the height of the sample in comparison with capillary rise can strongly affect experimental results.  Only for very tall samples, not initially filled completely via capillarity, are data independent of sample height.  This point is not widely appreciated in prior experiments, where sample heights tend to be small and are often not even reported, much less varied.  This point is also not widely appreciated in prior modeling efforts, none of which to our knowledge accounts for the influence of sample height.

The capillary bundle model we developed here has general significance for several reasons.  First, the sample height is now explicitly included in the analysis.  Second, it points to the importance of two key concepts not considered in prior work:  (1) The differential dimensionless expelled water curve, $E(P)$ versus $P$, defined by Eq.~(\ref{expwater}) as the incremental volume ${\rm d}V_w$ of water expelled per pressure increment ${\rm d}P$, made dimensionless by the sample cross sectional area $A_0$, the density $\rho$ of water, and $g=9.8$~m/s$^2$; and (2) The cumulative cross-sectional area fraction $f(r)$ of pores with radii less than $r$.  Whereas prior experiments focus on the volumetric water content and, prior modeling is in terms of the volumetric distribution of pore radii, the capillary bundle model shows how $E(P)$ is simply and directly related to $f(r)$ via
\begin{equation}
	 {\rho g \over A_0} { {\rm d}V_w \over {\rm d}P } = f\left(\frac{2\gamma}{P}\right)-f\left(\frac{2\gamma}{P+\rho gH_0}\right),
\label{model}
\end{equation}
where $\gamma$ is the liquid-air surface tension and $H_0$ is the sample height.  In words, the dimensionless differential volume of water expulsion per pressure increment is equal to the cross-sectional area fraction of active pores.   This expression makes explicit both the connection between pressure plate data and pore-scale structure, as well as the role of sample height.

In terms of our specific model-soil systems, we can make several conclusions based on our new data and its analysis via the capillary bundle model.  For randomly packed monodisperse spherical beads, the area distribution of the pore radii is well-approximated as a log-normal, peaking around one third the sphere radius and extending with full-width half-max between about $0.2-0.5$ sphere radii; the total cross-sectional area fraction of the pore space is found to be 0.34.  With the addition of superabsorbent hydrogel particle additives, this pore space is modified by the swelling of the gels.  The particle size is not crucial, provided it is not very much larger than the bead size.  Then only about 0.1 weight percent is required to clog up the original pores between beads and reduce the water-accessible pore space.  This can have a very significant effect on the permeability of the medium \cite{Verneuil11}.

There are several follow-up questions that would be of interest to the broader physics community.  What is the role of hydrogel particle shape and stiffness?  With regards to stiffness, to what extent can cohesion in real soils serve the same role as confinement in our lab experiments?  If soft and confined, then the swelling gels must conform to fixed pore space; if stiff and unconfined, then the swelling gels can unjam the soil.  The latter could lead to interesting dynamical effects, both regarding sample preparation protocol and also for cyclic wetting and drying.  At what level does the capillary bundle model break down, and the irregular geometry of the pore space matter?  Perhaps this could appear via hysteresis in retained water for cyclic variation of the pressure head.  Can our measurement of cross-sectional pore area distributions be confronted directly with simulation of sphere packings?   If the pore size is reduced, at fixed geometry, eventually water storage in wetting films must become important.  Can this be observed and understood by extension of the capillary bundle model?  Beyond these immediate questions, we hope that the general experimental and theoretical methods presented here may find future use for studying real soils in nature.   And we hope our work provides a general set of tools for reliable quantitative testing of trial additives designed to improve water usage efficiency in agriculture.

\begin{acknowledgments}
We thank Jean-Christophe Castaing and Zhiyun Chen in the Rhodia group of Solvay Inc.\ for helpful conversations.  We also thank Degussa Inc.\ for kindly providing the hydrogel particle samples.  This work is supported by the National Science Foundation through grants MRSEC/DMR-112090, DMR-0704147, and DMR-1305199.
\end{acknowledgments}


\bibliography{PressurePlate_References}


%
\end{document}